\documentstyle[epsfig]{aipproc}

\begin{document}
\title{New Millisecond Pulsars in Globular Clusters}

\author{Nichi D'Amico$^1$, Andrea Possenti$^1$, Richard N. Manchester$^2$, John Sarkissian$^3$, Andrew G. Lyne$^4$ and  Fernando Camilo$^5$}
\address{$^1$Osservatorio Astronomico di Bologna, via Ranzani 1, 
40127 Bologna, Italy}
\address{$^2$Australia Telescope National Facility,
CSIRO, PO Box 76, Epping, NSW 2121, Australia}
\address{$^3$Australia Telescope National Facility,
CSIRO, Parkes Observatory, PO Box 276, \\
Parkes, NSW 2870, Australia}
\address{$^4$University of Manchester, Jodrell Bank
Observatory, Macclesfield, SK11~9DL, UK}
\address{$^5$Columbia Astrophysics
Laboratory, Columbia University, 550 West 120th Street, \\
New York, 
NY 10027}

%\lefthead{LEFT head}
%\righthead{RIGHT head}
\maketitle

\begin{abstract}
A new search of globular clusters for millisecond pulsars is in
progress at Parkes. In this paper we describe the motivation, the
new hardware and software systems adopted, the survey plan and the
preliminary results. So far, we have discovered ten new millisecond
 pulsars in four clusters for which no associated pulsars were
previously known.
\end{abstract}

\section*{Introduction}
Exchange interactions in the core of globular clusters
result in the formation of binary systems containing neutron stars. In these
systems, the neutron star is eventually spun up through mass
accretion from the evolving 
companion \cite{sb76,bv91,ka96}, resulting  in the formation
of  millisecond pulsars.  These objects are
valuable for studies of the dynamics of clusters, the evolution of binaries, 
and the interstellar medium \cite{phi92,hmg+92,bv91b,fcl+00}.  But searches for millisecond pulsars are
difficult because they are usually rather weak and  their signals 
are distorted by propagation through the interstellar medium, and because 
the apparent spin period 
may be affected by doppler-shift changes due to binary motion.  

After several discoveries, made mainly in the early 1990s, no
additional  pulsars were found in globular clusters, leaving 
the question open why some clusters (e.g. 47
Tucanae or M15) had large numbers of detectable pulsars, whereas other 
apparently similar clusters have few or none.

When, a few years ago, a  new  multibeam 20-cm receiver was 
installed at Parkes, we decided to initiate a new search of  
globular clusters for millisecond pulsars.  This receiver has 
a system
temperature of $\sim 21$~K and bandwidth of $\sim$ 300 MHz, resulting in
an unprecented sensitivity. In order to further improve our 
search capability, we have constructed at Jodrell Bank and Bologna
a new high resolution
filterbank system, made of $512\times 0.5$ MHz adjacent pass-band filters.
This  makes possible to remove the effects of dispersion in 
the interstellar  medium more efficiently than previous searches,
and allows the  detection of 
millisecond pulsars with dispersion measures (DMs) of more than 
200 cm$^{-3}$pc. The combination of this new equipment with the relatively 
high frequency of the multibeam  receiver and its sensitivity level
gives a unique opportunity to probe distant  clusters. 
Also, because globular clusters are known to contain short-binary period 
millisecond pulsars, and because this class of objects is a very 
interesting one, we have implemented a new multi-dimensional code 
to search over a range of accelerations 
resulting from binary motion, in addition to the standard search  
over a range of
 DMs. 
 So far, we have discovered 10
millisecond pulsars in four clusters, none of which had previously known
pulsars associated with them. Five of these pulsars are members of
short-period binary systems, and four of them have relatively high DM values.

\section*{Observations and results}

We have selected about 60 clusters, on the basis of their central density
and distance. Observations consist usually of a 2.3h integration on each target.
The resulting nominal ($8\sigma$) sensitivity  to a
typical 3 ms pulsar with DM $\sim$ 200 cm$^{-3}$ pc is about 0.14 mJy,
several times better than previous searches.  Sampling the 512 
channels every 125 $\mu$s, 
each observation produces a  huge 
array, 32 Gsamples, or 4 Gbytes (packing the data at 1-bit/sample),
and requires significant CPU resources for offline processing.  
In Bologna, we have implemented the new code on a local cluster of 
Alpha-500MHz CPUs and on the Cray-T3E 256-processor system at 
the CINECA Supercomputing Center. 

In the off-line processing, each data stream
is split into non-overlapping segments of 2100, 4200 or 8400 sec and these
are separately processed.  When the DM is not known
precisely from the existence in a given cluster of a previously known
pulsar, data are first de-dispersed over a wide range of
dispersion measures centered on the value expected for each cluster on the
basis of a model of the Galactic electron layer \cite{tc93}, and
then transformed using a Fast Fourier Transform (FFT). 
The analyis method exploits
the fact that even relatively highly accelerated binaries might have significant 
spectral power in the FFT.  Time-domain data are fast-folded at
periods corresponding to a significant number of spectral features above 
a threshold to form a
series of `sub-integration arrays' and these arrays are searched for the
parabolic signatures of an accelerated periodicity.  Parameters for final
pulse profiles having significant signal-to-noise ratio are output for visual
examination. 

When a pulsar is detected and confirmed in a cluster, we usually reprocess
the data. The raw data are de-dispersed at the single DM value of the newly 
discovered pulsar; then the resulting time series is  interpolated 
to compensate for an acceleration and transformed using a FFT, with many 
trials to cover a large acceleration range. 
Since this
analysis involves many FFTs, it is relatively slow, and has benn
 rarely used when a  DM value (or a narrow DM range)  was not available.

\subsection*{Millisecond pulsars in NGC 6752}

NGC 6752 is believed to have a 
collapsed core and was already known to possess a large proportion of 
binary systems 
and dim X-ray sources.  In this cluster, we have first 
discovered a 3.26 ms pulsars in a 21~h orbital period binary system, 
PSR J1910$-$59A \cite{dlm+01}(see Fig 1). This pulsar has
a relatively low DM, 34 cm$^{-3}$pc and  scintillates markedly, similar to the 
pulsars in 47 Tucanae,  so it is seen rarely.  As has been already 
experienced on 47 Tucanae \cite{clf+00}, 
amplification due to scintillation might occasionally help in 
the detection 
of additional rather weak millisecond pulsars in the same cluster. 
And in fact, devoting a large amount of 
observing time to this cluster, we have already found four additional 
previously unseen millisecond pulsars  (Table 1).   
\begin{figure} [h]% fig 1
\centerline{\epsfig{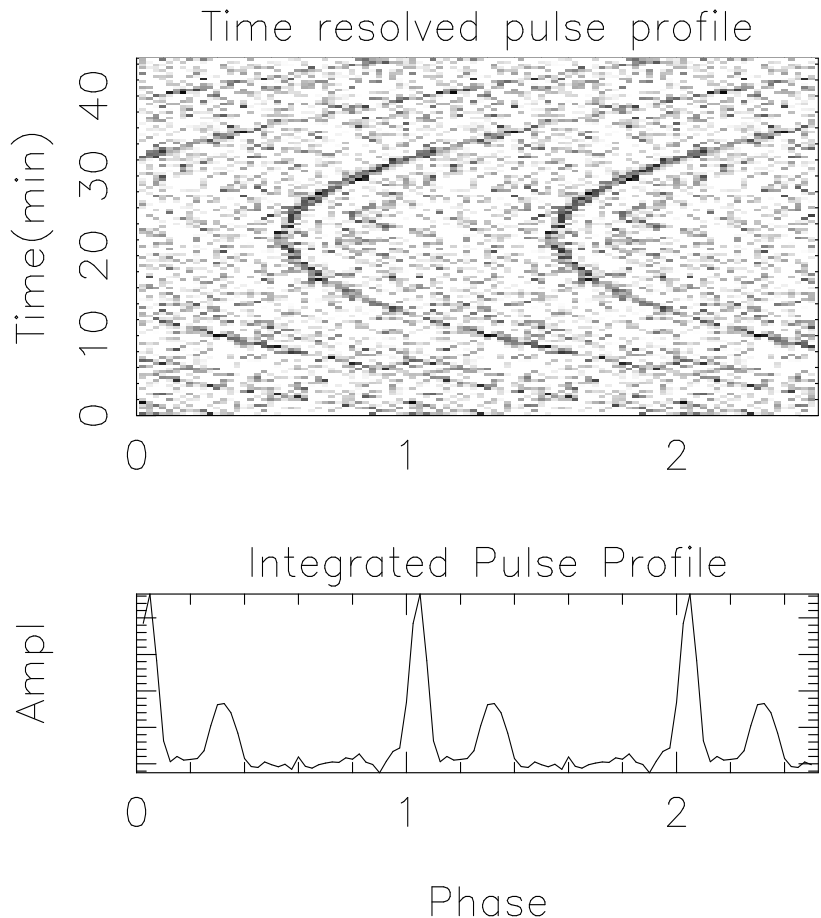}}
%%%%%%%%%%%\vspace{10pt}
\caption{Integrated (bottom panel) and time-resolved (top panel)
pulse profile for PSR J1910$-$58A in NGC6752.  The curvature in the
top panel show that the apparent period varies during the observation due
to the pulsar's orbital motion}
\label{fig1}
\end{figure}

\subsection*{An eclipsing millisecond pulsar in NGC 6397}
NGC 6397 is a prime candidate for globular cluster 
searches. It is close and has a very dense and probably collapsed core and it contains at least four X-ray sources,  but  there was no 
known pulsar associated with NGC 6397 prior to this search.  
In this 
cluster we
 have found PSR J1740$-$53,  a millisecond pulsar with a spin period of 
3.65ms and an orbital period of 1.35 days \cite{dlm+01}.  
This pulsar is eclipsed for more than
40 \% of the orbital phase. Similar eclipses are observed in other 
binary pulsars \cite{fbb+90,sbl+96}. But  these systems are close binary systems 
(with orbital periods of a few hours) and have relatively light 
companions (minimum mass $<$ 0.1 M$_{\odot}$). 
 In contrast, J1740-53 is in a rather wide binary system, with an orbital
period of 1.35 days, and has a heavier companion ($>$ 0.18 M$_{\odot}$).  
It seems unlikely that  a wind of sufficient density could be driven 
off a degenerate companion, and hence 
produce the observed eclipses.  Therefore, follow-up observations of this 
pulsar will be useful to probe the eclipse 
mechanism in millisecond pulsars.

\subsection*{Millisecond pulsars in NGC 6266}

We have discovered three millisecond binary pulsars 
in NGC 6266, another relatively dense cluster.  
The first one, PSR J1701$-$30A \cite{dlm+01}, has a spin 
period of 5.24 ms, an orbital period, 3.8 days, and the mass function 
indicates a minimum companion mass of 0.19 M$_{\odot}$.  
This system is similar to several low-mass binary pulsars, 
associated with globular clusters or in the Galactic field.  
But the two other systems found, PSR J1701$-$30B and PSR J1701$-$30C, 
belong to the class of short-binaries.  They have 
spin periods of 3.6 ms and 3.8 ms and  orbital periods of 3.8h and 
5.2h (Fig. 2).  
\begin{figure} [h]% fig 2
\centerline{\epsfig{file= 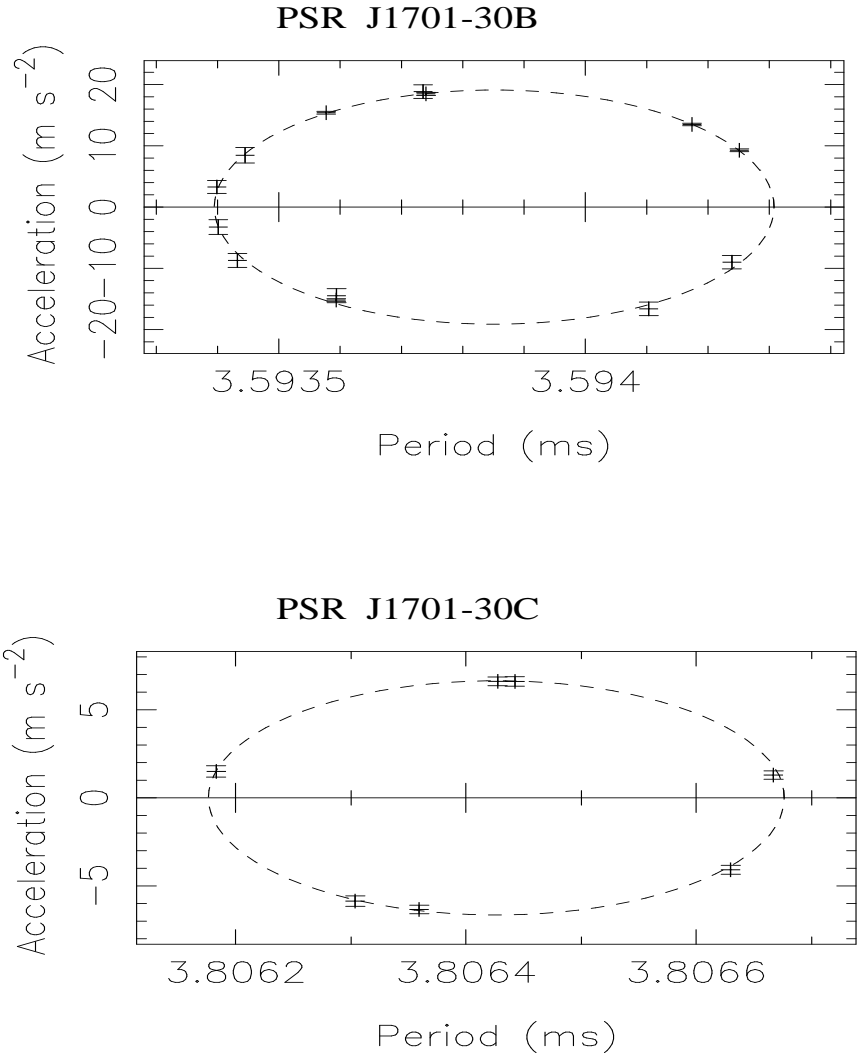,height=5.5in,width=4.5in}}
%%%%%%%%%%%\vspace{10pt}
\caption{Observed accelerations plotted against barycentric period for 
PSR J1701$-$30B and PSR J1701$-$30C in NGC 6266.  Dashed ellipses represent
best fits circular orbits given in Table 1}
\label{fig2}
\end{figure}

\subsection*{An ultra-short binary in  NGC 6544}

This cluster is also one of the closest,  and most concentrated 
globular clusters known.  The pulsar discovered, PSR J1807$-$24 
\cite{dlm+01,rans01}, has 
a spin period of 3.06 ms and it is binary, with an extremely short 
orbital period, 1.7 hours, 
the second shortest  known. Even more interestingly, the projected 
semi-major axis of the orbit is only
12 light-ms. The corresponding minimum companion mass is only 
0.0089 M$_{\odot}$ or about 10 Jupiter masses.  

\section*{Conclusions}

In Table 1 we report the preliminary parameters of the new millisecond
pulsars discovered so far in  four globular clusters. 
 It is too early to draw new  conclusions on the pulsar
content of globular cluster, as there are many clusters in our sample
that need to be searched. The
present experiment raises an interesting issue from the observational
point of view:  besides sensitivity and powerful search algorithms, 
a key strategy in a search for millisecond pulsars in globular clusters
is to devote a large amount of observing time to each target.  In fact, quite
often these objects have properties such that they can be seen very 
rarely only: 
scintillation in low DM 
clusters, abnormally long eclipses, and unfavourable  orbital phases 
in the case of ultra-short binaries might easily prevent the detection 
during a single observation. But the hidden systems are very
often the very interesting ones.

\begin{center}
\begin{small}
\begin{table}
\caption{Parameters of the millisecond pulsars discovered}
\label{table1}
\begin{tabular}{llllll} 
 Cluster & Pulsar & Period (ms) & DM (cm$^{-3}$ pc) & Orbital period (days) & Mass function (M$_\odot$) \\ \hline

 NGC6266 & J1701$-$30A & 5.2415660(4)   & 114.4(3) & 3.805(1) & 0.0031 \\
         & J1701$-$30B & 3.59386(2)     & 114.4(3) & 0.159(3) & 0.0009 \\
         & J1701$-$30C & 3.80643(2)     & 114.4(3) & 0.221(5) & 0.0002 \\ \hline

 NGC6397 & J1740$-$53  & 3.650328896926(9)    & 71.8(2)  & 1.35405971(2) & 0.0027 \\ \hline

 NGC6544 & J1807$-$24  & 3.0594487974(3)& 134.0(4) & 0.071092(1) & 3.85$\times$10$^{-7}$ \\ \hline

 NGC6752 & J1910$-$59A & 3.266186212(3)    & 34(1)    & 0.83711(1)    & 0.0029 \\
         & J1910$-$59B & 8.35779(1)     & 34(1)    &   -   &        -  \\
         & J1910$-$59C & 5.27732(2)     & 34(1)    &   -   &        -  \\
         & J1910$-$59D & 9.03528(2)     & 34(1)    &   -         &   -     \\
         & J1910$-$59E & 4.57177(2)     & 34(1)    &   -         &   -

\end{tabular}
\end{table}
\end{small}
\end{center}

\end{document}